# Adapting a Cryogenic Sapphire Oscillator for Very Long Baseline Interferometry

SHEPERD DOELEMAN [1], TAO MAI [1], ALAN E.E. ROGERS [1],
JOHN G. HARTNETT [2], MICHAEL E. TOBAR [2], AND NITIN NAND [2]

**ABSTRACT**. Extension of very long baseline interferometry (VLBI) to observing wavelengths shorter than 1.3mm provides exceptional angular resolution (~20 μas) and access to new spectral regimes for the study of astrophysical phenomena. To maintain phase coherence across a global VLBI array at these wavelengths requires that ultrastable frequency references be used for the heterodyne receivers at all participating telescopes. Hydrogen masers have traditionally been used as VLBI references, but atmospheric turbulence typically limits (sub) millimeter VLBI coherence times to ~1-30 s. Cryogenic Sapphire Oscillators (CSO) have better stability than Hydrogen masers on these time scale and are potential alternatives to masers as VLBI references. Here, We describe the design, implementation and tests of a system to produce a 10 MHz VLBI frequency standard from the microwave (11.2 GHz) output of a CSO. To improve long-term stability of the new reference, the CSO was locked to the timing signal from the Global Positioning System satellites and corrected for the oscillator aging. The long-term performance of the CSO was measured by comparison against a hydrogen maser in the same laboratory. The superb short-term performance, along with the improved long-term performance achieved by conditioning, makes the CSO a suitable reference for VLBI at wavelengths less than 1.3mm.

---

[1] MIT Haystack Observatory, Westford, Massachusetts, USA, 01886; sdoeleman@haystack.mit.edu, tm2419@columbia.edu, aeer@haystack.mit.edu
[2] University of Western Australia, Crawley, WA, Australia, 609; john@physics.uwa.edu.au, mike@physics.uwa.edu.au, nandn01@student.uwa.edu.au.

# 1. INTRODUCTION

Building on early pioneering efforts (Padin et al 1990; Greve et al. 1995; Krichbaum et al. 1998), very long baseline interferometry (VLBI) at wavelengths shorter than 3 mm has steadily improved in capability. A major challenge for such arrays is that most millimeter and submillimeter radio telescopes have much smaller diameters than their centimeter wavelength counterparts, which limits VLBI sensitivity. However, development of modern digital VLBI back ends and hard-disk based high-speed recorders has dramatically increased VLBI bandwidths, compensating for the relatively smaller dish sizes to some extent. Now, ad hoc VLBI arrays of global millimeter and submillimeter facilities can detect and resolve many interesting astrophysical targets with angular resolutions of a few tens of microarcseconds.

Of particular interest is Sagittarius A* (SgrA*), a compact radio source associated with the $4 \times 10^6$ solar mass black hole candidate at the center of the Milky Way (Reid 2009 and references therein). Because of its proximity (~8 kpc) and mass, Sgr A* presents us with the largest apparent event horizon of any black hole candidate in the universe and the possibility of directly observing signatures of general relativity in the strong gravity regime. Recent 1.3 mm VLBI observations have detected structure within Sgr A* that is on the same scale (~5 Schwarzschild radii) as theoretical predictions that take into account extreme gravitational lensing effects on radiation near the event horizon (Doeleman et al. 2008). As VLBI array sensitivity at short wavelengths increases it will be possible to resolve time-variable structures near the black hole event horizon that are predicted by accretion theory (Doeleman et al. 2009a; Fish et al. 2009). VLBI at short wavelengths is essential for study of this object since scattering in the interstellar medium, which depends on wavelength as $\lambda^2$, blurs VLBI images at longer wavelengths and obscures the intrinsic structure.

To increase the resolution of observations of Sgr A* and other targets requires extension of VLBI arrays to even shorter (<0.8mm) wavelengths. In this regime, turbulence in the atmosphere limits the time interval over which the VLBI signals can be coherently integrated. Typical coherence times at these wavelengths are between 1 and 30 s. On these timescales and at wavelengths of 0.8 mm or less, the stability of typical hydrogen masers begins to introduce coherence losses. Ideally, one would employ a frequency reference with better stability than a hydrogen maser on these short timescales, but that also has the maser's excellent long-term stability (>10,000 s) characteristics. Since the geometry of a VLBI array is never perfectly known, it is standard practice to search for VLBI detections over residual delay and fringe frequency. Minimizing the long-term delay and frequency drifts of VLBI references allows this search space for detection to be narrowed, significantly increasing the efficiency of VLBI data processing.

Our group has been exploring alternatives to hydrogen masers for this short-wavelength VLBI work. Cryogenic sapphire oscillators (CSOs) have frequency stability that could be an order of magnitude better than hydrogen masers at timescales shorter than 10 s (Doeleman 2009b; Hartnett et al. 2006) and have the potential to serve as VLBI frequency standards for submillimeter wavelength work. CSOs are prone to long-term frequency drift (Tobar et al. 2006), which degrades their stability on longer timescales; however, recent developments using cryorefrigerators have resulted in CSOs with very small frequency drift (Hartnett & Nand, 2010; Grop et al. 2010). Costa et al (1992) reported the first 2.3GHz and 10GHz VLBI observations that used a CSO as the fundamental frequency reference. Costa et al (1992) found that use of the CSO for VLBI was superior to using a rubidium gas cell, but at these relatively low observing frequencies, use of hydrogen masers as VLBI references incurs negligible coherence losses, and there is no compelling scientific rationale to replace them with a CSO. Other groups have developed CSO frequency standards for deep space communication applications (Wang et al. 2004, 2005) and have also employed CSO-stabilized references for VLBI observations at centimeter wavelengths.

This work describes a CSO-based VLBI frequency standard for use at submillimeter observing wavelengths where the superior stability of the CSO from 1-100 s integration times reduces coherence losses due to hydrogen maser standards. A CSO developed at the University of Western Australia (Hartnett et al. 2009) was installed for development and testing at the MIT Haystack Observatory Westford Radio Telescope site. The native CSO output frequency, 11.2 GHz, is too high to be conveniently used as a VLBI reference, and requires down-conversion to a VLBI-standard 10 MHz tone. A specialized receiver that phase-locks a 10MHz quartz oscillator to the CSO was designed to generate a station VLBI reference. In order to increase the long-term stability of the 10 MHz output, a feedback control loop is introduced in the CSO receiver to condition the 10 MHz output to the reference signal received from GPS satellites, which is highly stable on timescales longer than a day. The control loop is implemented in software using a proportional-integral controller mechanism. Multiple parameters of the control loop need to be optimized according to various factors in the system. Since the control loop is designed to improve long-term stability, it has a time constant that is typically a few days long, which makes it very time-consuming to determine the parameter values by trials on the real system. An efficient solution was to use a software simulation to determine the optimum loop parameters. In addition, we have conducted experiments to determine the sensitivity of the CSO to temperature and magnetic field, which provides information on how to improve the CSO stability by compensating for ambient changes.

## 2.0 RECEIVER TO DERIVE 10 MHz FROM THE CSO

In order to make a cryogenic sapphire oscillator useful for VLBI we need to derive a 10 MHz signal from the CSO with optimum stability and minimum drift relative to the national frequency standard broadcast by the GPS. The CSO outputs a frequency near 11.2 GHz that has superb short- and medium-term stability but that degrades on timescales longer than about $10^3$ s.

A block diagram of the receiver is shown in Figure 1. A high-quality 10 MHz crystal oscillator is phase-locked to the CSO. This crystal provides very low phase noise at frequencies more than 30 Hz from the carrier achieving values of -155, -165, and -165 dBc $Hz^{-1}$ at $10^2$, $10^3$, and $10^4$ Hz from the carrier respectively.

The phase lock is implemented by synthesizing a signal at 11.140 GHz from the 10 MHz, which is mixed with the CSO to form an intermediate frequency (IF) of 60 MHz. The IF is then mixed with the output from a direct digital synthesizer (DDS) to provide the phase difference between the CSO and the signals synthesized from the crystal. The phase detector is then filtered using an analog loop filter. This loop filter has a loop bandwidth of 30 Hz and a damping constant of 0.7 based on a phase detector sensitivity of 0.1 V rad$^{-1}$ at 11 GHz and a crystal oscillator tuning sensitivity of 0.4 Hz V$^{-1}$ at 10 MHz or 440 Hz V$^{-1}$ at 11 GHz. The DDS also provides a quadrature output, which provides an indication of lock and a measure of detector sensitivity and signal level when phase detected.

The DDS has a frequency resolution of 1 μHz, which allows a very fine fractional frequency adjustment to bring the 10 MHz derived from the CSO into alignment with GPS on a timescales longer than about $10^5$ s. This is accomplished by comparing the CSO 10 MHz with 10 MHz from a Trimble Thunderbolt GPS-disciplined oscillator.

The parameters of the GPS phase-lock loop are under software control. The phase between the CSO and GPS is derived from a quadrature phase detector from which an unambiguous phase is derived using the C atan2 function. The GPS loop makes the CSO track GPS on a long timescale by changing the DDS. A change of one in the least significant digit of the DDS corresponds to a change of one part in the $10^{16}$ in the 10 MHz output.

## 2.1. Performance of the Receiver

The receiver was tested with the GPS loop turned off, substituting an Agilent E8257D synthesizer for the CSO and comparing the 10 MHz output of the receiver with the 10 MHz output of the Agilent synthesizer using a Symmetricom TSC5115A phase-noise test set. Table 1 gives the measured residual phase-noise performance of the CSO receiver using the SLSM3 synthesizer, as shown in Figure 1, as well as the phase-noise performance when an Agilent E8257D-UNX is substituted for the SLSM3.

The CSO receiver performance is limited, in the short term, by the phase noise of the synthesizer and in the longer term by the change of synthesizer phase with temperature. The Luff Research SLSM3 synthesizer had a temperature coefficient of $-6 \pm 2\,ps\,K^{-1}$. The Agilent synthesizer temperature coefficient was measured by comparing two synthesizers, using a common reference, and changing the temperature of one of the synthesizers. Temperature coefficients of $-6 \pm 2\,ps\,K^{-1}$ and $-9 \pm 3\,ps\,K^{-1}$ were measured for the two units.

TABLE 1

CSO RECEIVER PHASE NOISE USING TWO DIFFERENT SYNTHESIZERS

| Freq. offset at 10 MHz | Luff (dBc Hz⁻¹) | E8257D-UNX (dBc Hz⁻¹) |
|:---:|:---:|:---:|
| 1 Hz | -120 | -130 |
| 10 Hz | -130 | -142 |
| $10^2$ Hz | -145 | -145 |
| $10^3$ Hz | -150 | -150 |

Table 2 gives the measured Allan standard deviations for the E8257D and SLSM3 synthesizers and for the noise floor for the TSC5115A phase-noise test set used to make the measurements.

TABLE 2

CSO RECEIVER STABILITY USING TWO DIFFERENT SYNTHESIZERS

| | Allan standard deviation | | |
|---|---|---|---|
| $\tau$ (s) | E8257D-UNX | SLSM3 | TSC 5115A[a] |
| $10^{-2}$ | | $8\times10^{-12}$ | $3\times10^{-12}$ |
| $10^{-1}$ | $8\times10^{-13}$ | $1\times10^{-12}$ | $3\times10^{-13}$ |
| $10^{0}$ | $8\times10^{-14}$ | $1\times10^{-13}$ | $4\times10^{-14}$ |
| $10^{1}$ | $1\times10^{-14}$ | $1.5\times10^{-14}$ | $6\times10^{-15}$ |
| $10^{2}$ | $1\times10^{-15}$ | $1.5\times10^{-15}$ | $6\times10^{-16}$ |
| $10^{3}$ | $3\times10^{-16}$ | $3.0\times10^{-16}$ | $1\times10^{-16}$ |
| $10^{4}$ | $\cdots$ | $1\times10^{-16}$ | $\cdots$ |

[a]TSC 5115A test equipment limits are shown

The temperature sensitivity limits the performance. For example if the ambient temperature changes by 1˚C in 500 s, the temperature coefficient of 10 ps K$^{-1}$ limits performance to about $2\times10^{-14}$.

## 2.2. The GPS Control Loop

The output of the CSO receiver and the 10 MHz GPS clock signal are fed into a quadrature phase detector and an unambiguous phase from the phase detector output is derived in software using the **atan2** function. The GPS control loop is implemented as a C program running on the computer that controls the CSO receiver. The program steers the CSO receiver output by adjusting the frequency of the DDS every 10s, a period that is short enough for the purpose of the control loop. A proportional-integral (PI) controller with pre-low-pass filtering of the GPS phase is implemented in the GPS control loop. The block diagram is shown in Figure 2. The 10 MHz from the GPS receiver is orders of magnitude noisier than the output of the CSO receiver on time-scale less than $10^3$ s, and the GPS signal is subject to diurnal variations of the ionosphere. To prevent the noisy GPS signal from degrading the performance of the CSO on short timescales, the phase detector output is averaged over time in the control loop. The averaging process can be approximated as a first-order low-pass filter with a time constant equal to the averaging time. The averaging introduces a phase shift in the system, and the averaging time needs to be much less than the loop time constant to avoid making the loop unstable. From the block diagram shown in Figure 2, we obtain the transfer function of the feedback loop as the ratio of the CSO phase to the GPS phase:

$$\frac{\phi_{cso}}{\phi_{gps}} = \frac{Ps+1}{s^2 + Ps + I + s^3\,\tau_{avg}\big/2\pi}$$

where $P$ is the proportional gain, $I$ is the integral gain, and $\tau_{avg}$ is the GPS averaging time.

The natural frequency $\omega_n$ and the damping coefficient $\zeta$ of the loop are $\omega_n = \sqrt{I}$ and $\zeta = P\big/\big(2\sqrt{I}\big)$. In the controller program, the loop time constant $\tau$ and the damping factor $\zeta$, instead of the proportional and integral gain, are directly specified by the user. The proportional and integral gains are derived accordingly as $P = 4\pi\zeta / \tau$ and $I = 4\pi^2 / \tau^2$ respectively.

It can be shown that choosing $\tau > (5/\zeta)\tau_{avg}$ ensures a phase margin better than $45°$. Although the actual averaging process is different from first-order low-pass filtering, we found through simulation that making $\tau_{avg} \sim \tau/7$ was about optimal. As long as the averaging time is small compared to the time constant, the system is approximately a second-order PI control loop. The averaging time for the phase error, loop time constant, and damping factor are the three key parameters of the control loop that need to be optimized. In addition to the PI feedback control, we introduced feedforward control into the system to compensate for aging of the CSO and the frequency shift due to temperature change. Given that we know the effect of temperature on the output frequency and the aging rate of the CSO, we can apply a correction in the opposite direction that cancels the frequency shift. The performance of the feedforward temperature compensation depends on accurate temperature coefficient and time constant of the response to temperature changes, so we have conducted experiments to determine these parameters with some success.

## 3. FACTORS CONTRIBUTING TO THE CSO INSTABILITY

### 3.1. Noise of the CSO

A plot of the Allan deviation of the CSO fractional frequency is shown in Figure 3. The plot of CSO vs. CSO2 is relatively flat for averaging times τ < 100 s, while on longer timescales, the slope starts to increase, most likely as a result of a random-walk of frequency and a constant frequency drift. The plot of the CSO without drift appears to have a slope τ½ for averaging times τ > 1000 s, which matches the characteristics of random-walk frequency noise. Therefore, for the simulations we assumed that the CSO output is dominated by flicker frequency noise and random-walk frequency noise on timescales over 1000 s, which is the range in which our GPS feedback loop operates. The 10 MHz output of the CSO receiver includes noise from the CSO and noise from the two synthesizers. The CSO receiver performance is limited by the Luff SLSM3 synthesizer, the performance of which was measured in the laboratory. We can see that the Allan deviation of the Luff synthesizer fractional frequency falls below $10^{-15}$ for averaging times longer than a few hundred seconds, while the CSO, as shown in Figure 3, has an Allan deviation above $10^{-15}$ for averaging times τ > 100 s, which continues to rise for longer averaging times. Therefore, the output of the CSO receiver on timescales greater than 1000 s is dominated by noise from the CSO itself as a combination of flicker frequency noise, random-walk frequency noise, and a constant frequency drift due to aging.

### 3.2. Aging of the CSO

The CSO under investigation continues to age indefinitely. Figure 4 shows the fractional frequency deviation of the 10 MHz output of the CSO receiver, which was not locked to GPS, but compared with the H maser (NR04) at Westford. By fitting a straight line to the plot, we calculated the aging rate, in terms of fractional frequency, to be about $1.42 \times 10^{-13}$ day$^{-1}$. Other variations seen in the plot are mainly due to changes in the room temperature.

### 3.3. Temperature Sensitivity of CSO Controller

The CSO frequency is observed to change with the temperature of the electronics, which maintains the oscillations in the sapphire crystal and is located outside the dewar. Temperature also changes the delay in the receiver electronics, which in turn affects the frequency. However, the effect of delay change is minor compared to the frequency shift of the controller directly caused by temperature change. We have measured the temperature dependence of the CSO oscillator control box by heating it up with an incandescent light. The light was left on for about 6 hr and then turned off, to approximate a step change in ambient temperature at the controller box. The response of the CSO output frequency is shown in Figure 5. We obtained a temperature coefficient of about $-2.1 \times 10^{-13} \mathrm{K}^{-1}$ and time constant of about 1.5 hr by fitting an exponential curve to the plot. And these numbers were used in the controller program to see how well model the change in the CSO rate with temperature can be compensated in the software. The CSO was operated in a room where temperature is regulated by a heater and an air conditioner to keep the daily temperature fluctuation to within 1 $^{\circ}$C. Without compensation, this led to a daily fractional frequency variation of about $2 \times 10^{-13}$.

### 3.4. Effect of Magnetic Field on the CSO Frequency

Since the frequency standard used for VLBI is often located close to the antenna, which can alter the local magnetic field when it rotates, we tested the sensitivity of the CSO by installing a coil of 20 turns of wire around the dewar of the CSO so that the sapphire crystal was at the center of the coil. Current was sent through the coil to generate a magnetic field that has the same order of magnitude as Earth's magnetic field. With this experimental setup, we could only perturb the ambient field magnetic in the vertical direction. The CSO was observed to be sensitive to a downward field perturbation. Assuming the frequency shift is proportional to the magnetic field intensity, we calculated the sensitivity to be about $-1.1 \times 10^{-13}$ $\mathrm{G}^{-1}$. Similar experiments have been done on another CSO to test its sensitivity (Wolf et al. 2004). In that experiment, a sinusoidal vertical magnetic field of 0.1 G was used, and a sensitivity of about $7 \times 10^{-15}$ $\mathrm{G}^{-1}$ was measured, which is 2 orders of magnitude smaller than what we measured at the Westford site. A possible explanation of the sensitivity to magnetic field, suggested by Kovacich et al. (1997) is a mode-frequency shift of a sapphire dielectric resonator that results from the particular paramagnetic impurities present in the crystal, their concentration, the polarization of the resonant mode, and how far the resonant mode frequency is from the dominant electron spin resonance line. For example, a recent experiment (Benmessai et al. 2009) measured the response to an axial DC magnetic field of a whispering gallery mode in sapphire at the $Fe^{3+}$ spin resonance of 12.04 GHz. The maximum measured sensitivity was determined to be $4 \times 10^{-10} \mathrm{G}^{-1}$ for fields of the order of 15 G. However, for fields below 1 G the sensitivity was asymmetric: $10^{-10} \mathrm{G}^{-1}$ for positive fields and near zero for negative fields. The variation of magnetic field due to the antenna dish movement should not exceed about 0.5 G, which corresponds to a frequency shift of less than $5 \times 10^{-14}$. Such magnetic sensitivity should not be a problem at sites where the CSO is located far from the antenna and any other magnetic disturbances. So far we have not employed any mechanism to correct frequency shifts due to magnetic field variation, assuming that its effect is minor compared to other sources of instability.

# 4. GPS REFERENCE SIGNAL

The GPS reference used to condition the CSO receiver is generated by a Trimble Thunderbolt GPS-disciplined clock, which outputs a 10 MHz signal derived from the L1 signal from GPS satellites. According to the data sheet of the GPS receiver, the Allan deviation of the 10 MHz output is about $10^{-10}$ for $\tau = 10$s and less than $10^{-12}$ for $\tau > 10^4$ s. The ionosphere also causes a significant delay in the signal from the GPS satellite. As the ionosphere changes significantly during a day, the phase of the GPS signal can change by tens of nanoseconds. The noise from the GPS receiver itself is relatively small compared with the diurnal variation due to the ionosphere. Based on observation results of the Millstone Hill radar, the total electron count (TEC) in the ionosphere above Haystack Observatory has a daily variation that ranges from a few TEC units (1 TECU = $10^{16}$ $e$ m$^{-2}$) to a few tens of TECU at different times of the year. A variation of 20 TECU corresponds to a change in delay of about 22ns at the GPS L1 frequency of 1575.42 MHz, for a signal received from the zenith. The delay will also change as the elevation angle of the satellite changes. As a result, a significant diurnal variation is seen in the 10 MHz output of the GPS receiver. Proper averaging of the phase error in the control loop is necessary to filter out such diurnal changes.

## 4.1. GPS Control Loop Simulations

Parameters of the GPS feedback loop need to be set properly to optimize the performance of the controller. As discussed previously, the three parameters are the averaging time for the phase error, the nominal loop time constant and damping coefficient. Given the long loop time needed to condition the CSO with GPS, it takes days for the system to settle, which makes it extremely time-consuming to determine the parameters by experimenting on the real system. To solve the problem more efficiently, a software simulation program was written to simulate the operation of the real system. The simulation program allows us to see the result of a week-long experiment in a few seconds. It also allows experiments with arbitrary input signals. For example, the step response of the system can be seen by feeding in a step function as the input to the simulation.

The simulation program operates in the same way as the real GPS feedback control program. It uses the same step size, 10 s, and the same formulas to calculate the frequency correction and temperature compensation. The only difference is that in the simulation, all the inputs to the feedback loop are generated in software. The inputs include the 10 MHz signals from the GPS receiver and the CSO receives, as well as the ambient temperature measured at the receiver. A CSO receiver, both locked and unlocked to GPS, were simulated in the program so that we could compare their performances. The output of the simulation program includes time-domain plots of both of the locked and unlocked CSO receiver outputs and the Allan deviation plots of both CSO signals and the GPS signal. In addition to using a randomly generated signal, the program provides an option of reading in the CSO receiver signal and the temperature from external data files, which allows us to run the simulation with real data. By using real data from the CSO not locked to GPS, the performance of the feedback loop can be evaluated more realistically in the simulation. To show that the simulation models the actual system realistically, plots are shown in Figures 6 and 7. Figure 6 shows the ringing of the actual system after a sudden change of the loop parameters. Figure 7 shows the simulated response of the system using a step function as input. The overshoot, decay rate, and period of oscillation of both plots closely match each other. GPS diurnal variation is set to zero in the simulation for a cleaner result.

## 4.2. Modeling Inputs to the Feedback Loop

To make the simulation realistic, the various inputs to the feedback loop need to be modeled correctly. The three inputs are the output of the free-running CSO receiver, the reference from the GPS receiver, and the ambient temperature. The simulation models the CSO receiver output as a signal with flicker frequency noise, a constant frequency drift, and sensitivity to temperature. We did not include random-walk frequency noise in the simulation because it has an effect similar to a constant frequency drift on long-term stability. For flicker frequency noise, the spectral power density of frequency fluctuation, $Sy(f)$, is proportional to $1/f$. In the simulation, the noise signal is generated by first constructing the signal in the frequency domain as a Gaussian random variable whose standard deviation is proportional to $1/f^{-1/2}$ and then transforming it to the time domain using an inverse fast Fourier transform. The frequency drift rate is set as the actual aging rate as measured. Since we lack a better frequency standard as reference to measure noise from the CSO on short timescales, the magnitude of the flicker noise is determined by matching the Allan deviation of the simulation result to that of the real data, the 100 s average frequency count of the CSO receiver measured by the TSC5115A. When using a noise signal with a magnitude of about $7 \times 10^{-15}$ rms, in fractional frequency, the Allan deviation for $\tau = 100$ s and $\tau = 10^5$ s of the simulated unlocked CSO differs from that of the real data by less than 5%. The resulting plots are shown in Figures 8a and 8b using real data as input and in Figures 9a and 9b using simulated data as input. In the simulation, the temperature sensitivity of the CSO is turned off to eliminate the effect of temperature. On the Allan deviation plot with real data, we can see a bump that ranges from about $3 \times 10^3$ s to $5 \times 10^4$ s on the plot of the unlocked CSO, caused by diurnal variations of temperature. But diurnal temperature variation should not affect system stability on scales of 100 s or $10^5$ s. The Allan Deviation rises with a constant slope on longer timescales because of aging of the sapphire. We can see that the Allan Deviation of the simulated unlocked CSO signal closely matches that of the real data. The flicker frequency noise used in the simulation is about $7 \times 10^{-13}$ rms.

The diurnal temperature variation is simulated as a simple sinusoid, which may not be realistic enough, but it is sufficient for showing the effect of temperature variation on system stability. Although the simulation allows temperature compensation, the result may not be meaningful because we can only use an ideal model for the response to temperature, which provides little information on the performance of temperature compensation on the real system. The signal from the GPS receiver is modeled as a signal with white phase noise and a sinusoidal diurnal variation in the phase delay. We were not aware of the significant diurnal variation of GPS before we closed the GPS loop on the real system. By matching simulation result to real data, we found that the variation of the phase delay of $\sim 10$ ns peak to peak. Although the actual GPS delay is not sinusoidal, the model is good enough for our purpose.

## 4.3. Optimizing Loop Parameters

The Allan deviation of a GPS signal with no diurnal variations and pure white phase noise falls of as $1/\tau$, where $\tau$ is the averaging time. With white phase noise of about 3.8 ns, the Allan deviation plot of the GPS signal will cross that of the unlocked CSO at $\tau \sim 5 \times 10^4$. The GPS becomes more stable than the CSO on timescales longer than the crossing point. So it was reasonable to set the GPS loop time constant to about $10^5$ s and to set the phase error averaging time to $10^4$ s. Increasing phase error averaging time helps to remove noise from the GPS, but the averaging time has an upper bound of about 15% of the loop time constant, due to system stability requirements. After activating the GPS loop on the CSO receiver on day 175, we observed a diurnal variation in the output frequency of $10^{-12}$ peak to peak, which corresponds to a variation in time delay of ~ 9 ns peak to peak. (This matches calculations done earlier using electron counts of the ionosphere.) The frequency is shown in Figure 10. Such fluctuation was too big to be a result of temperature changes given the temperature coefficient we had measured. So it is most likely result of GPS. On day 178, the error averaging time was increased to $3 \times 10^4$ s, and the loop time constant was raised to $3 \times 10^5$ s to ensure stability. We can see that tripling the averaging time reduces the fluctuation by a factor of about 3.

To eliminate the effect of the GPS diurnal variation, the phase error needs to be set to almost one day or higher. We used an averaging time of 1 day (86,400 s). Simulations with different values for the damping coefficient and the time constant showed that a damping factor of 0.8 and a time constant of $6 \times 10^5$ produce favorable response. Using a larger time constant would unnecessarily slow down the feedback loop. Using a smaller time constant, on the other hand, would prolong the settling time of the system, and the system would be unable to stabilize when the time constant became smaller than $3.3 \times 10^5$ s. Figures 11 and 12 show the result of a simulation using the parameters stated previously and without temperature compensation. We can see that the Allan deviation of the GPS-disciplined CSO is almost half a decade lower than that of the free-running CSO, and it drops below $10^{-14}$ for averaging time over $10^6$ s (about 11.6 days). Due to the diurnal variations of GPS, it is not practical to correct frequency shift due to diurnal temperature variation by correcting with GPS so we have to rely entirely on feedforward control for temperature compensation. Figure 13 shows the measured Allan deviation of the CSO locked to GPS with a loop time constant of 6.9 days. For these data, which were taken from day 221 through day 226, there was no feedforward correction of the temperature applied, because it was determined that the overall temperature coefficient of the CSO was less than one part in $10^{13}$, presumably because the coefficient of the oscillator control box is compensated by a coefficient of the opposite sign from the dewar or another part of the CSO. The H maser being used as a reference also changes rate with temperature. The lack of perfect temperature control of the enclosure in which the maser is housed results in some diurnal variation at the level of a few parts in $10^{14}$.

## 5. FUTURE DEVELOPMENT

So far we have collected only limited data to show improvement of the long-term stability of the CSO when conditioned with GPS. More data are needed in the future to determine the performance of the CSO receiver on timescales longer than a few days. Experiments are also needed to analyze the effectiveness of temperature compensation.

Currently the ambient temperature is measured by a sensor in the CSO receiver that is a few meters away from the CSO controller box. Although the CSO receiver box has good air ventilation, the temperature inside the receiver is higher than the room temperature and may be affected by operations of the receiver electronics. The temperature measured may not accurately indicate the ambient temperature changes at the CSO controller box, which degrades the performance of temperature feedforward control. If we can measure the temperature  of the sensitive parts of the CSO directly with sensors, the effectiveness of temperature compensation should improve. Further improvement would result from placing the CSO in a more constant-temperature environment.

Another major limitation of the feedback control system is the strong diurnal variation of the GPS signal, which forces us to use large error averaging time and loop time constant. As a result, the feedback loop becomes really slow, and frequency variations within a day cannot be corrected by the feedback controller. The current GPS receiver only receives the L1 signal, and provides no correction for the ionospheric delay. GPS receivers that capture both L1 and L2 signals are able to automatically correct the ionospheric delay and provide a much more stable reference. A better GPS reference would allow us to lock the CSO receiver more tightly on the GPS by using smaller time constant and shorter error averaging time. With a wider bandwidth, the feedback loop would even be able to help reduce frequency shifts due to diurnal temperature variation. The magnetic field sensitivity of the CSO has only been investigated using weak fields in the vertical direction, and further experimentation for a more detailed characterization would be useful.

In addition the short-term stability of the frequency synthesis in the CSO receiver needs further development to realize the full potential of the CSO on times scales less than about 10 s.

## 6. VLBI COHERENCE

The VLBI coherence can be estimated from the Allan standard deviation using equation (12) of Rogers and Moran (1981):

$$\left\langle C^2\left(T\right)\right\rangle^{\frac{1}{2}} = \left[\frac{2}{T}\int_0^T\left\{\exp\left[-\frac{\omega^2\tau^2}{4}\left(\sigma_y^2\left(\tau\right)+\sigma_y^2\left(2\tau\right)+\cdots\right)\right]\right\}\left(1-\tau/T\right)d\tau\right]^{\frac{1}{2}},$$

where $T$ is the coherent integration time (in seconds), $\omega$ is the observing frequency (in rad s$^{-1}$), and $\sigma_y^2\left(\tau\right)$ is the Allan variance at averaging time $\tau$

We have computed the coherence losses for VLBI baselines at the 0.8 mm observing wavelength using this expression for several cases, and results are plotted in Figure 14. For each baseline we assume either a specific frequency reference at each end of the baseline (with no coherence loss due to atmospheric stability is specified (with no coherence loss due to imperfect frequency references). Table 3 summarizes the fractional stability of the references (labeled a through d) and of the assumed atmospheres (labeled A through D). For the coherence calculation, the Allan standard deviation at intermediate averaging times are interpolated from the values given in the table. In Table 3, the Allan deviation values for the H maser are for a state-of-the-art model, while the values for the CSO with no drift removed, the CSO conditioned by GPS, and the crystal are results from the measurements of each reference compared against a H maser at the Haystack Observatory Westford site. To derive a conservative estimate of fractional stability, the stability of the state-of-the-art H maser was subtracted in quadrature from these comparison measurements. The atmospheric stability above the ALMA (Atacama Large Millimeter Array) site (case B) comes from measurements in Nikolic (2009). The ALMA-corrected atmosphere is the projected residual atmospheric atmospheric stability at the ALMA after correction by water vapor radiometers (WVRs) and assumes a 50 μm rms phase out to 100 s, followed by a factor-of-10 decrease with every decade increase in averaging time beyond 100 s (Nikolic 2009). The Jet Propulsion Laboratory (JPL)-corrected atmosphere is based on the WVR-corrected measurements of Naudet et al (2000) on a 13 km baseline at the Goldstone site in Mojave, California. The wet summer atmosphere is typical for sites at low elevation in the summer months.

TABLE 3

<small>STABILITY COMPARISON OF VARIOUS FREQUENCY STANDARDS AND ATMOSPHERIC CONDITIONS</small>

| Case | Allan standard deviation | | | | | Case description |
|------|------|------|------|------|------|------|
| | 1 s | 10 s | 100 s | 1000 s | 10,000 s | |
| a | $8\times10^{-14}$ | $1.5\times10^{-14}$ | $3\times10^{-15}$ | $8\times10^{-16}$ | $4\times10^{-16}$ | H-maser |
| b | $8\times10^{-14}$ | $5\times10^{-15}$ | $2\times10^{-15}$ | $2\times10^{-15}$ | $1\times10^{-14}$ | CSO-GPS |
| c | $8\times10^{-14}$ | $5\times10^{-15}$ | $1.5\times10^{-15}$ | $4\times10^{-15}$ | $3\times10^{-14}$ | CSO-no drift correction |
| d | $8\times10^{-14}$ | $8\times10^{-14}$ | $1\times10^{-13}$ | $1\times10^{-13}$ | $1\times10^{-13}$ | Crystal 8607 |
| A | $1.5\times10^{-15}$ | $1.5\times10^{-15}$ | $1.5\times10^{-15}$ | $1.5\times10^{-16}$ | $1.5\times10^{-17}$ | Alma Corrected |
| B | $1.3\times10^{-14}$ | $1.3\times10^{-14}$ | $1.3\times10^{-14}$ | $1.3\times10^{-15}$ | $1.3\times10^{-16}$ | Alma Uncorrected |
| C | $1\times10^{-13}$ | $1\times10^{-13}$ | $2\times10^{-14}$ | $3\times10^{-15}$ | $2\times10^{-15}$ | JPL corrected |
| D | $1\times10^{-12}$ | $1\times10^{-12}$ | $1\times10^{-12}$ | $1\times10^{-13}$ | $1\times10^{-14}$ | Wet summer |

# 7. VLBI TEST

Once the Allan deviation of the CSO reference was measured in the laboratory, we conducted a VLBI observation to verify the operational performance of the new standard. On 2010 July 8, the quasar 4C39.25 was tracked for 10 minutes using a single VLBI baseline linking the Haystack Observatory Westford Antenna (18m) in Westford, Massachusetts with the Goddard Geophysical and Astronomical Observatory (GGAO) 6m antenna in Greenbelt, Maryland. A net bandwidth of 256 MHz, centered at 8668 MHz was recorded at each site using Digital VLBI back ends and Mark5B VLBI recorders. With Nyquist sampling at two bits, the aggregate data rate for a single polarization was 1Gbit s$^{-1}$. At GGAO, a hydrogen maser provided the frequency reference for all VLBI instrumentation. At Westford, three separate and identical VLBI systems were configured to record the observed passband, each using a different frequency reference for comparison. One reference was a standard hydrogen maser used for all geodetic VLBI observations at Westford, the second was the GPS-conditioned CSO described previously, and the third was a high-stability quartz crystal oscillator made by Oscilloquartz (OCXO model no. 8607).

At this observing frequency, the advantages of using the CSO over the hydrogen maser are negligible, and even the quartz crystal performs nearly as well. Figure 15 shows the VLBI phases determined by cross-correlating recorded signals using each of the different frequency references. Correlations among the three signals recorded at the Westford site confirm that both the Westford hydrogen maser and the CSO are both more stable than the crystal oscillator. From the VLBI data, for example, we find that over 1 minute integration times the Allan deviation is $6.6\times10^{-14}$ for the CSO-crystal pair, $6.2\times10^{-14}$ for the maser-crystal pair, but only $2.8\times10^{-14}$ for the maser-CSO pair (Fig. 15, left). On this 800 km VLBI baseline, and at this observing frequency, the atmosphere dominates the phase fluctuations on all integration times, and the coherence of the VLBI data is independent of which frequency reference is used (Fig 15, right). Given the excellent short-term stability of the CSO as confirmed through the comparisons described in § 4, this successful field test indicates that the CSO can provide a low coherence loss reference for future high-frequency VLBI observations.

## 8. SUMMARY

Cryogenic sapphire oscillators have superior short-term stability and could potentially be better frequency standards than hydrogen masers for VLBI at frequencies of 345 GHz and higher. But they are prone to frequency drifting on long timescales. A control system has been developed to improve the long-term stability of the CSO at the MIT Haystack Observatory Westford site by locking it to the GPS reference signal and by compensating for frequency shift due to ambient temperature changes. Software simulation tools were developed to help determine the optimal values for system parameters, and to predict long-term system performance. Experiments have also been performed to measure various factors that affect the stabilities of the system, which include the CSO's sensitivity to temperature and magnetic field and the noise and diurnal variations of the GPS signal. The long term performance of the system is currently limited mostly by the variations in the GPS signal due to ionospheric delay. Improvements can be made in the future by using a two-channel GPS receiver that corrects for ionospheric delay, and by keeping the CSO at a constant-temperature environment. With the GPS feedback control system, the CSO's long-term stability could be improved to the same level as a hydrogen maser or even better. We also note recent improvements in the long term performance of CSOs cooled with cryocoolers (Hartnett & Nand 2010).

We thank David Fields, Mike Poirier, and Bruce Whittier for technical assistance with the Cryogenic Sapphire Oscillator. We also thank Chris Beaudoin, Arthur Niell, Chester Ruszczyk, and Michael Titus for assistance with the very long baseline interferometry (VLBI) observations. The VLBI data presented here were collected in parallel with a test carried out by the NASA Geodetic VLBI program, and we gratefully acknowledge their support. This work was supported through NSF grant AST-0722168.

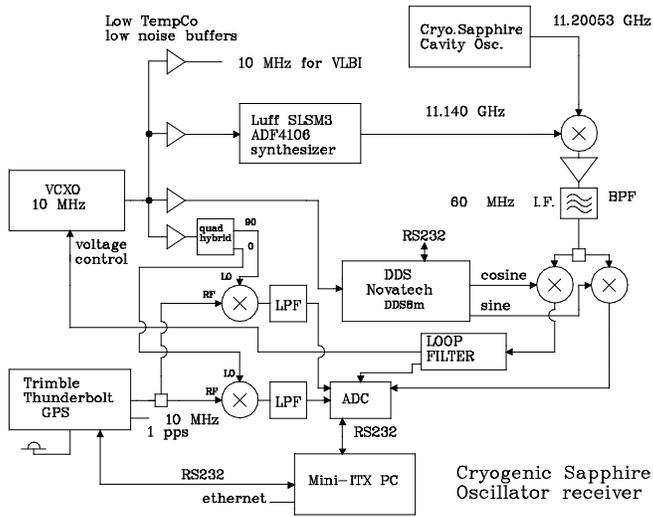

Figure 1. Block diagram of the CSO receiver.

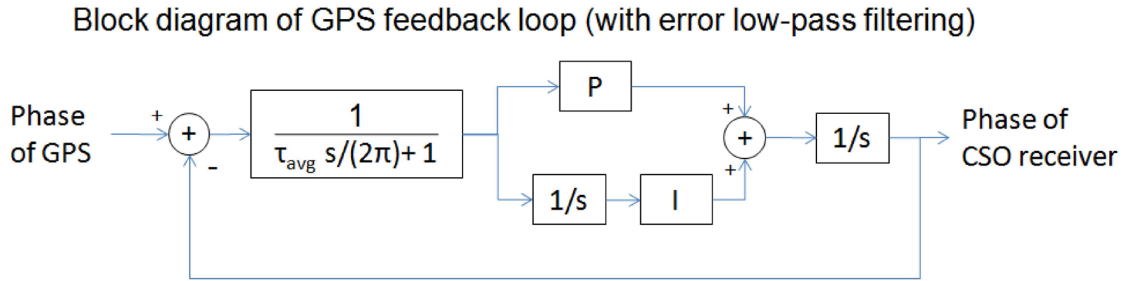

Figure 2. Block diagram of GPS control loop with low-pass filtering of error.

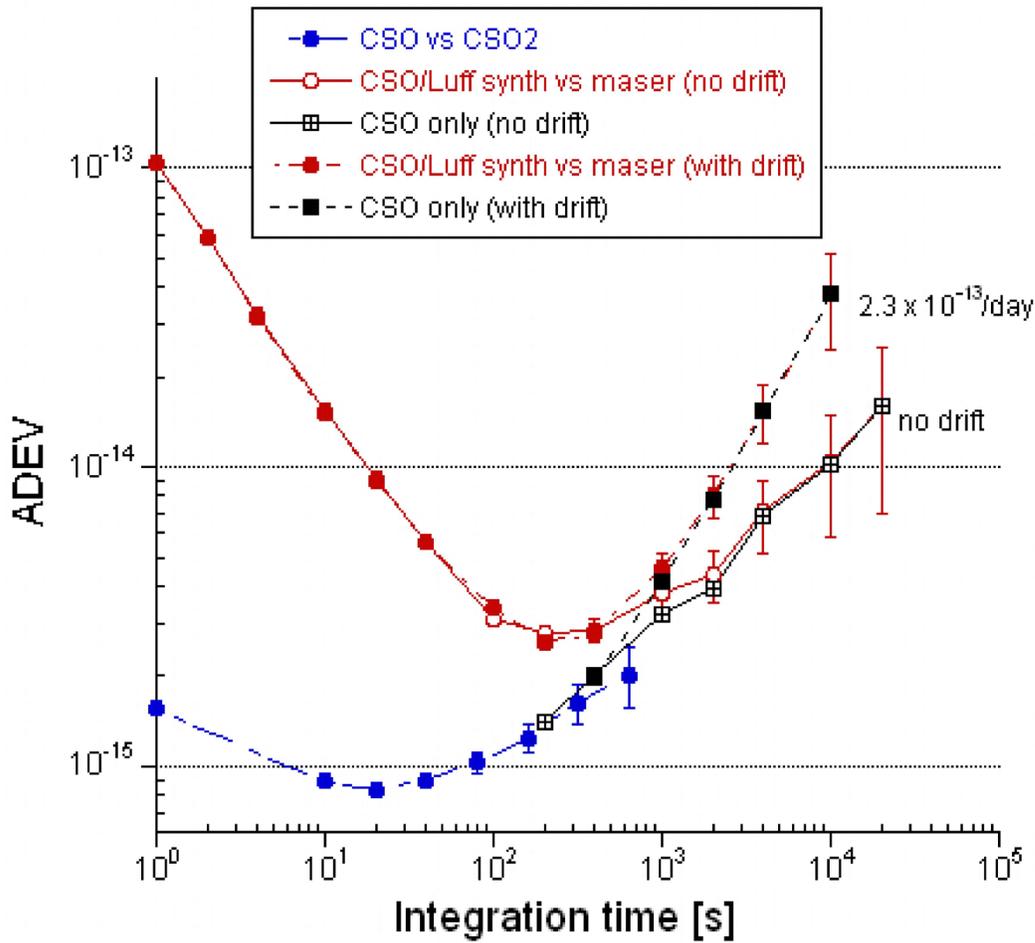

Figure 3. Allan deviation of fractional frequency of the CSO with various references. CSO vs CSO2 curve is from a comparison made in the University of Western Australia lab where two similar CSO were compared. CSO/Luff synthesizer vs maser curves, with and without frequency drift, were made at Haystack using a H maser as the reference; hence the short-term stability is dominated by the maser stability. The long-term stability of the CSO-only curves was estimated from the latter, with and without frequency drift. The fractional frequency drift was estimated as $2.3 \times 10^{-13} \, \text{day}^{-1}$ soon after the CSO was commissioned.

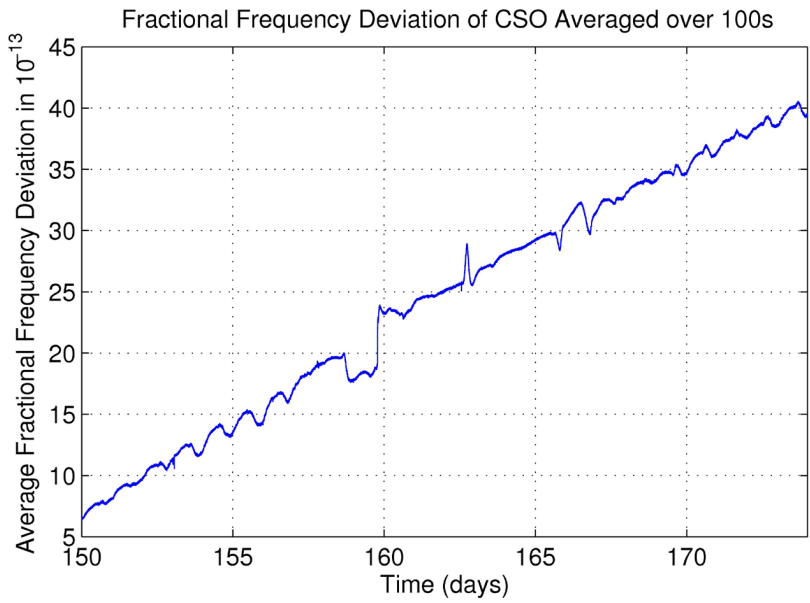

Figure 4. Frequency drift of CSO receiver due to aging of CSO itself.

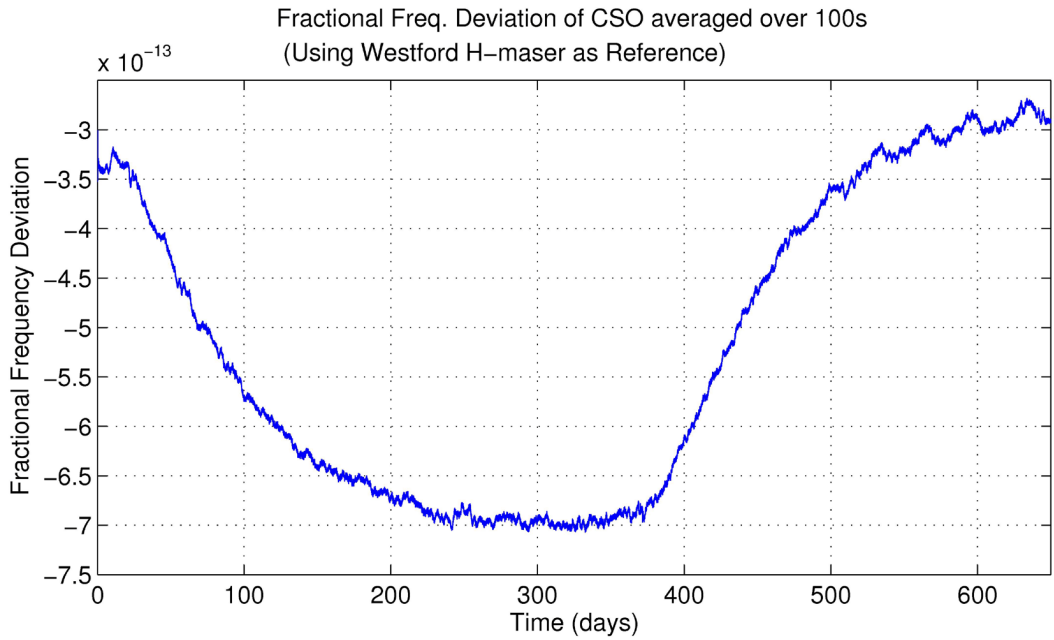

Figure 5. Response of CSO frequency due to step change in ambient temperature.

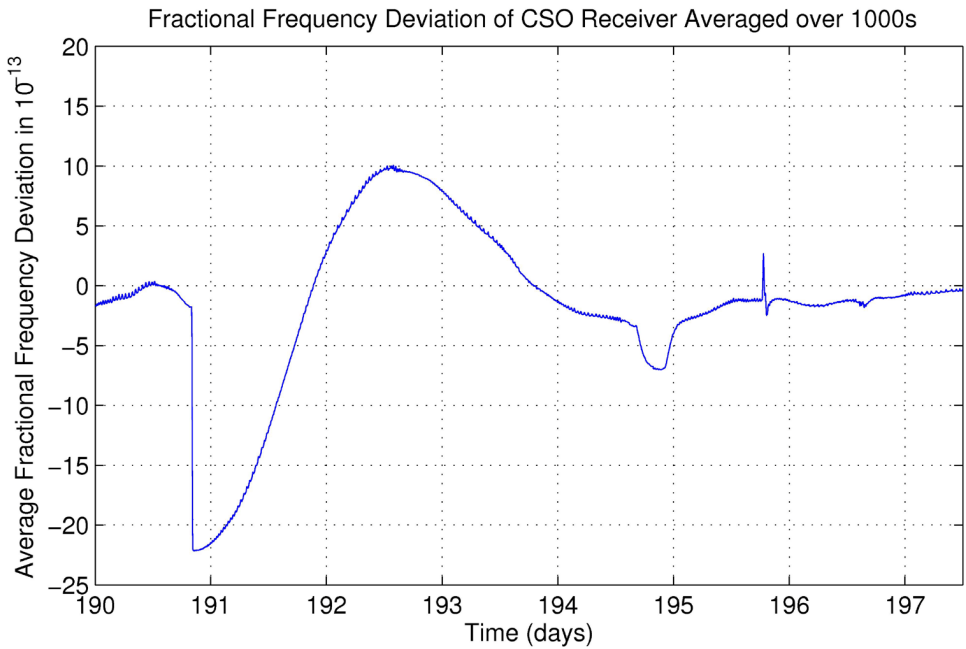

Figure 6. Step response of the GPS-disciplined system. The small frequency fluctuations in the last third of day 194 and the last quarter of day 195 are caused by ambient temperature changes, and are not part of the step response of the system.

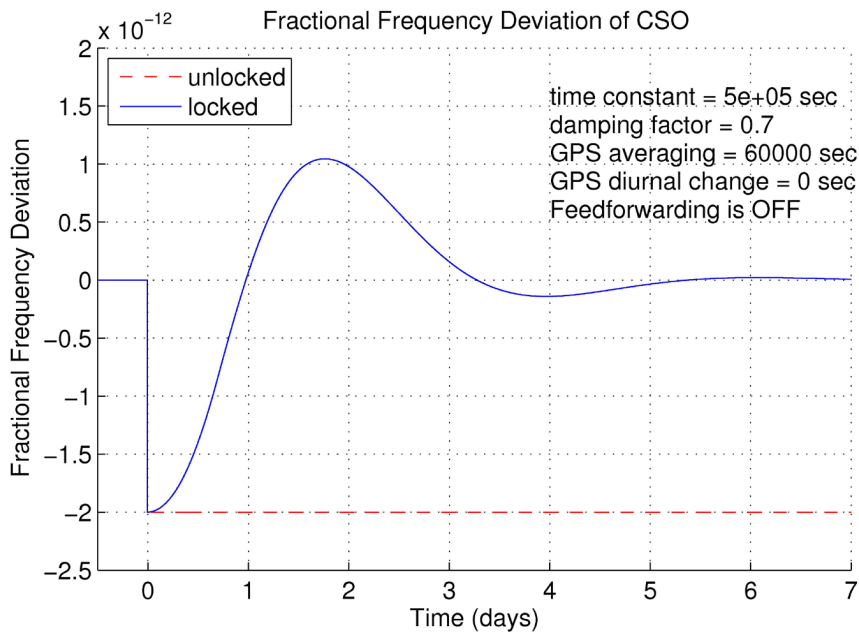

Figure 7. Step response of the feedback-controlled system in simulation closely matches the response of the actual system.

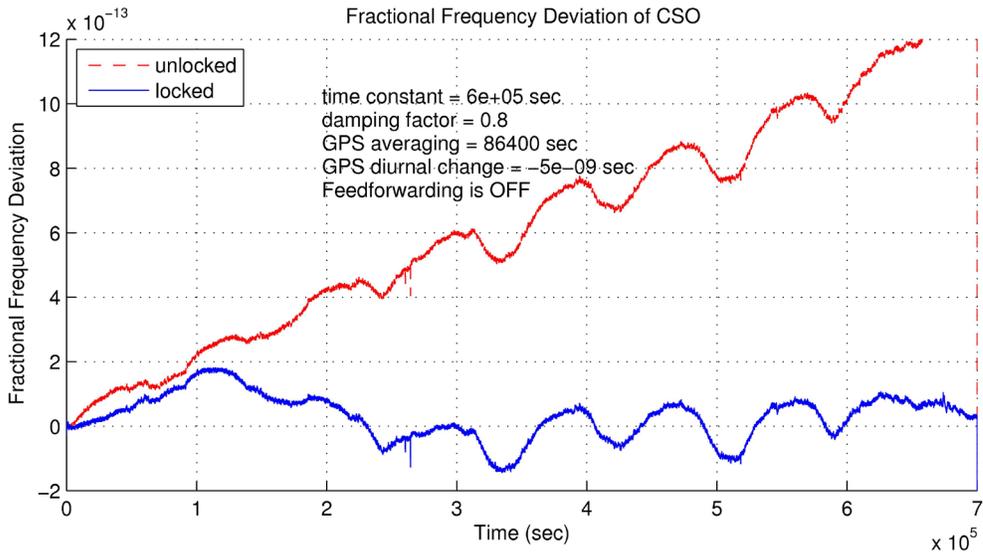

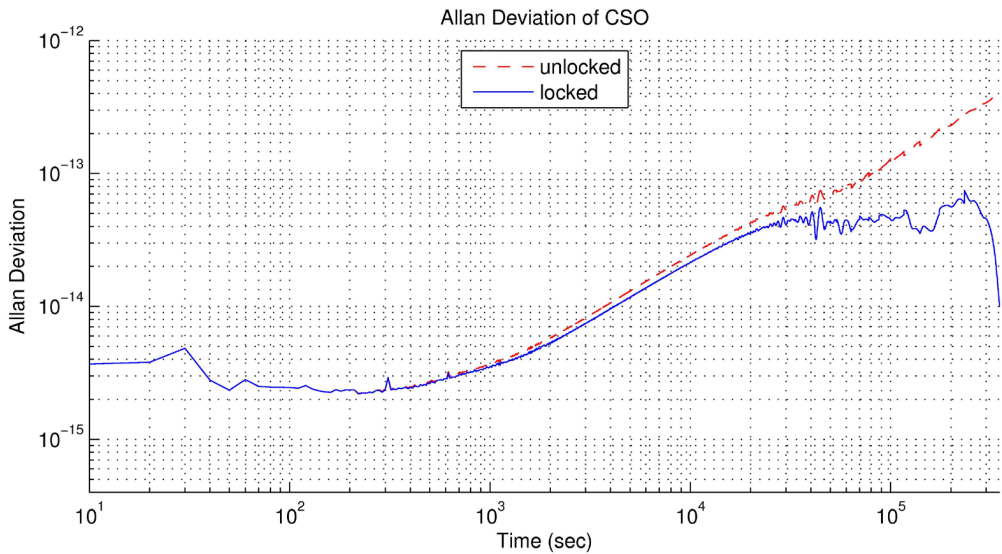

Figure 8. Simulation of the GPS-disciplined system (locked) using actual output of the free running system (unlocked). The data for the free running system is collected by the TSC 5115A and that for the controlled system is obtained from simulation.

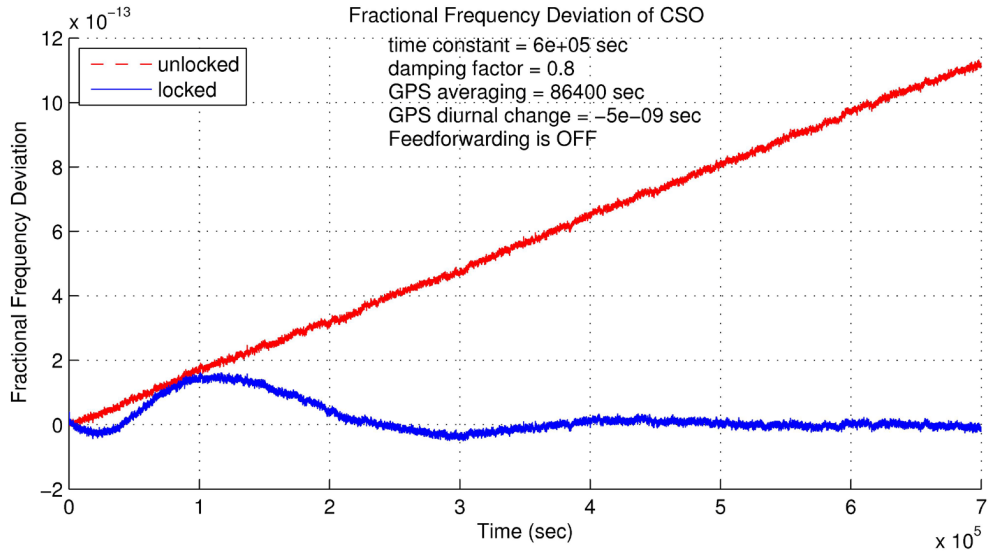

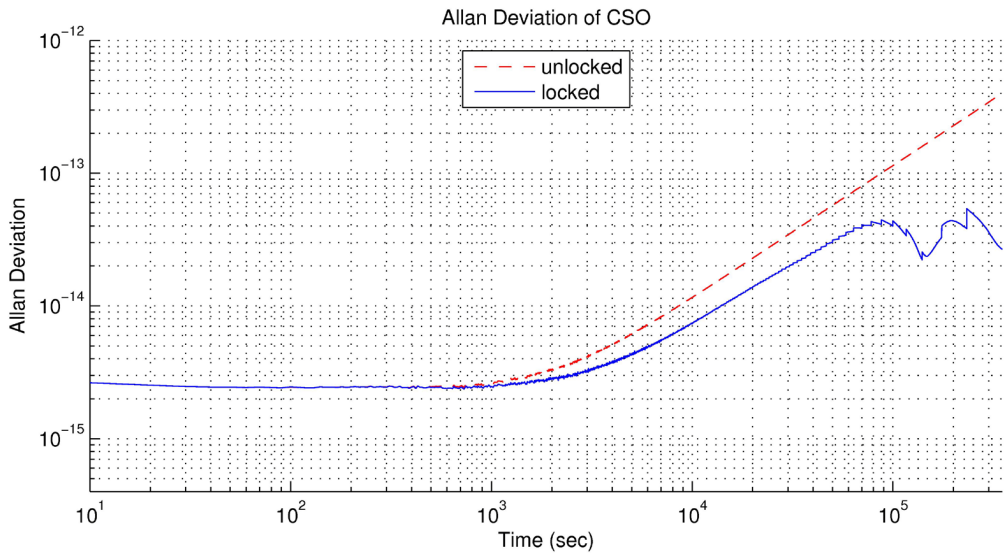

Figure 9. Simulation using software generated data. Note that the Allan deviation of the simulated undisciplined (unlocked) CSO closely matches that of the actual undisciplined CSO, as shown in Fig. 8. The flicker frequency noise used in the simulation is about $7 \times 10^{-13}$ rms.

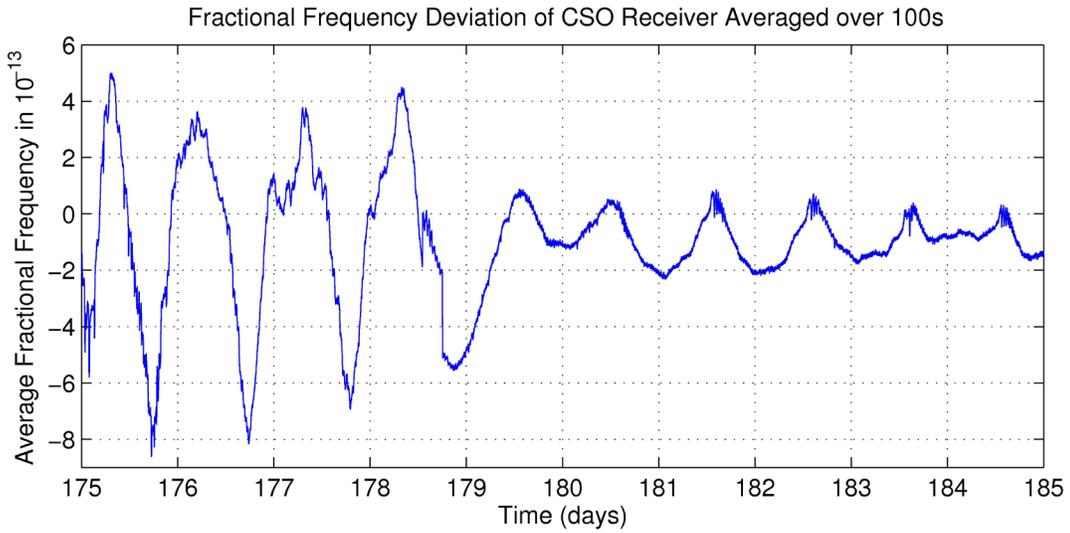

Figure 10. Frequency fluctuation of the GPS-disciplined CSO receiver due to diurnal variations of the GPS signal.

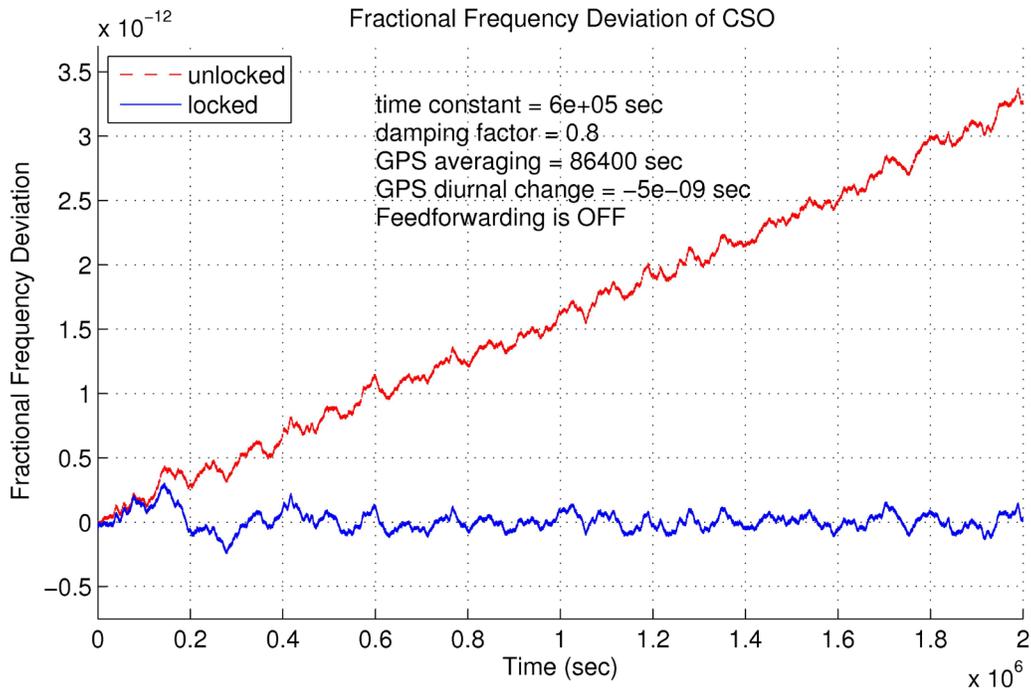

Figure 11.  Simulation with loop time constant 6 x10⁵ s, damping factor 0.8,  GPS averaging time 86400 s, and no feedforward.

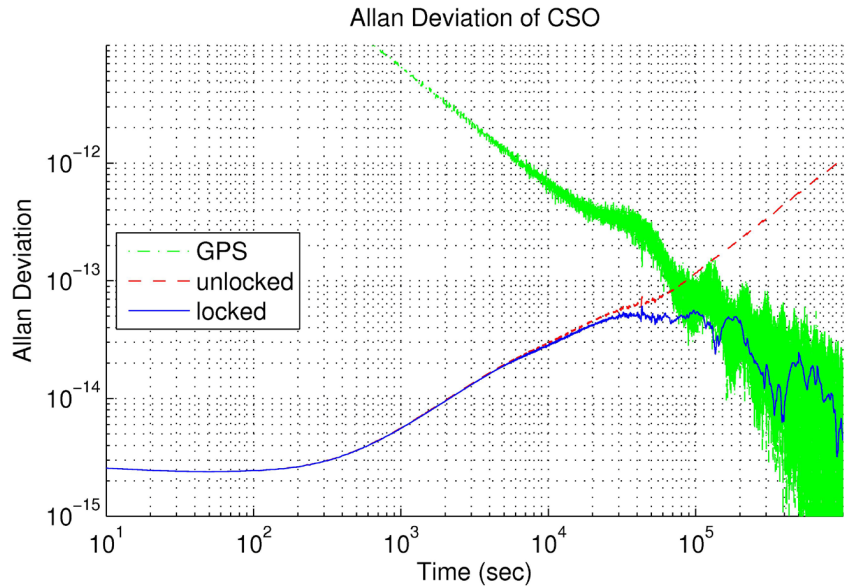

Figure 12. Allan deviations of the GPS-disciplined (locked) CSO, the un-disciplined (unlocked) CSO, and the GPS reference signal from the simulation described in Figure 11.

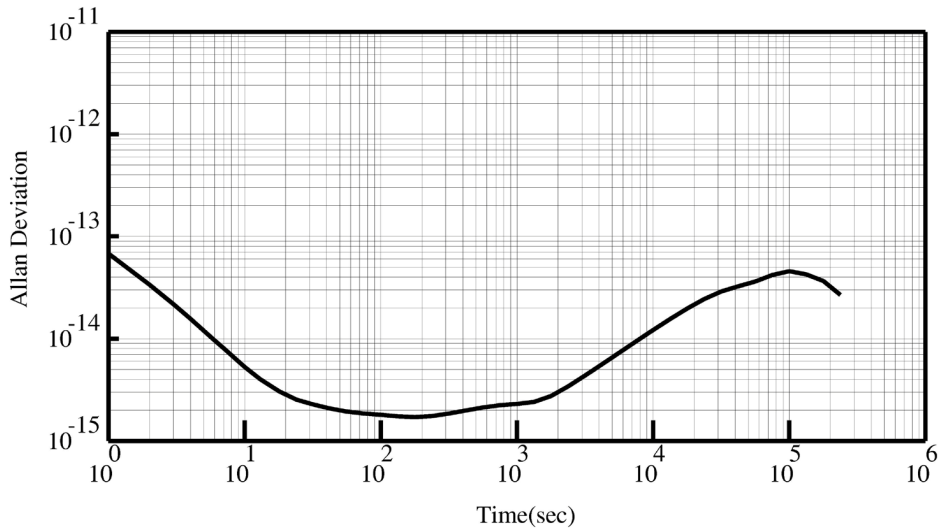

Figure 13. Measured performance of CSO locked to GPS with pre-low pass filter time constant of 86000s, loop time constant of 6 x $10^5$ s, loop damping of 0.8 and feed-forward compensation of CSO aging rate of 1.42 x $10^{-13}$ per day. Measurments used a H-maser as reference and have been corrected for the assumed values of Allan deviation of the H-maser given in Table 3.

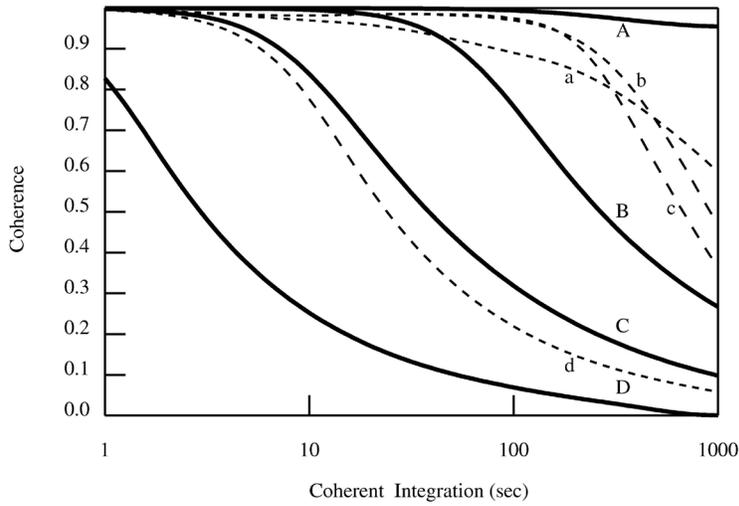

Figure 14. Estimated VLBI coherence at 345 GHz for the various cases listed in Table 3. The thin lines are for the frequency standards alone, assuming the same standard at each site, and the thick lines are for the atmospheres alone, assuming the same atmosphere at each site.

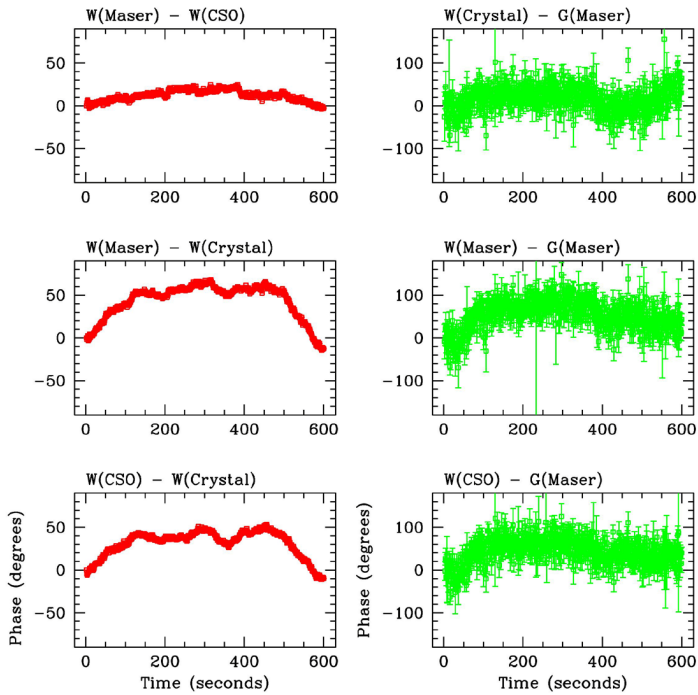

Figure 15. Results of using the CSO as a frequency reference for a VLBI observation between the Haystack Observatory Westford site and the GGAO antenna in Greenbelt, MD. At Westford, three separate VLBI systems used three different frequency references: the CSO, a Hydrogen maser, and a quartz crystal. At GGAO, a Hydrogen maser was used. Left panels show phases derived from cross-correlating the three receiver signals at the Westford site. The quartz crystal shows more pronounced phase variations compared to the two other references. Panels on the right show phases determined from cross-correlating all three signals recorded at Westford against the signal recorded at GGAO. On the long VLBI baseline, atmospheric effects dominate the phase fluctuations (note the scale difference between the left and right plots).